\documentclass[review]{elsarticle}

\usepackage{times}
\usepackage{w-thm}
\usepackage{comment}
\usepackage{xcolor}
\usepackage[]{graphicx}

\def\grad{\nabla}

\def\1{\perp 1}
\def\2{\perp 2}

\def\beq{\begin{equation}}
\def\eeq{\end{equation}}
\def\beqar{\begin{eqnarray}}
\def\eeqar{\end{eqnarray}}

\def\grad{\nabla}

\newcommand\usecolor{black}

\begin{document}

\begin{frontmatter}
  \title{Machine Learning approach to modeling of \\  neutral particles transport in plasma}

  % Author 1 is at address 1
  \author[addr1]{M.V. Umansky\corref{cor1}}
  \ead{umansky1@llnl.gov}
    
  % Author 2 is at address 1 and 2
  \author[addr2]{G.J. Parker}
    
  % Author 3 is at address 3
  \author[addr3]{R.D. Smirnov}
  %\ead[url]{www.researcher.com}
    
  % Corresponding author footnote
  \cortext[cor1]{Corresponding author}
    
  % Author footnote
  %\fntext[fn2]{Another author footnote}
    
  % Address definitions
  \address[addr1]{Lawrence Livermore National Laboratory, Livermore, CA 94550, USA}
  \address[addr2]{University of California, Berkeley, Department of Mathematics, Berkeley CA 94720, USA}
  \address[addr3]{University of California San Diego, Department of Mechanical and Aerospace Engineering, La Jolla, CA 92093-0411, USA}

  \begin{abstract}
    A propagator-based approach is investigated for Monte-Carlo (MC)
    modeling of neutral particles transport in fusion boundary
    plasmas. The propagator is based on a Green’s function for the
    neutral kinetic equation, which depends on the plasma profiles. A
    Neural Network (NN)-based model for the propagator provides a fast
    and accurate solution for the neutral distribution function in
    plasma. Preliminary results from a small 1D test problem look
    encouraging. The proposed approach, a propagator-based NN model for
    neutral transport in plasma, has potential for generalization to
    higher dimensions and efficient coupling with plasma models.
  \end{abstract}


\end{frontmatter}

%%%%%%%%%%%%%%%%%%%%%%%%%%%%%%%%%%%%%%%%%%%%%%%%%%%%%%%%%%%%%%%%%%%%%%%%%%%
\section{Introduction}

Detailed modeling of boundary plasma physics is necessary for
designing a practical magnetic fusion device. However a plasma model
alone is insufficient to address many problems in the edge plasma
where the neutral particles strongly influence the plasma; thus a
neutral model is needed to allow for self-consistent plasma-neutral
calculations \cite{Heifetz1986}.

Neutral particles are present in the edge plasma of magnetic fusion
devices due to interactions between the plasma and the wall and
divertor materials, as well as due to beam injection and neutral gas
puffs. These neutrals interact with plasma particles through processes
such as electron-impact ionization, charge exchange, and radiative
recombination. These collisional processes control neutral transport
into the core and affect plasma dynamics. They also play a role in
plasma fueling via recycling, whereby ions impinging on the wall are
re-emitted as neutrals and then become ionized. Furthermore,
interaction of high-energy neutrals with material wall may result in
sputtering of material surfaces and production of impurity ions in the
plasma.

\textcolor{\usecolor}{
  In many present-day experiments,
}
fusion boundary plasmas are rather collisional, as measured by
$\lambda_{e,i} \ll L_{||}$ and $\rho_{e,i} \ll L_{\perp}$, where
$\lambda_{e,i}$ is the electron and ion Coulomb collisional
mean-free-path (MFP), $\rho_{e,i}$ is the gyro-radius, and $L_{||}$,
$L_{\perp}$ are the parallel and perpendicular scale
length. Therefore, fluid models for fusion boundary plasmas remain
actively used. For neutral particles in plasma, fluid models are
applicable, when the neutral MFP, $\lambda_N$, is small compared to
the characteristic spatial scale lengths, $L_{||}$, $L_{\perp}$, and
fluid neutrals models are used in some edge plasma simulation codes
\cite{Rognlien1992,Blommaert2018}. However, the assumption of short neutral MFP
often breaks down in the edge of existing magnetic fusion devices, and
it probably will not hold, at least not universally, in future
devices.

Kinetic models for neutrals are necessary to accurately capture both
long and short MFP neutrals, and in the context of fluid edge plasma
modeling, the treatment of neutrals has traditionally been handled
with kinetic Monte Carlo (MC) codes \cite{Heifetz1986}.  However, MC
neutral models are subject to statistical noise, and coupling such a
model with a continuum-based plasma model for integrated
plasma-neutral simulation one usually finds that the neutral
statistical noise can interfere with the accuracy and convergence of
the coupled model \cite{Ghoos2016}.

Several continuum-based (non MC) kinetic codes for neutrals in plasma
have been developed in the recent years \cite{Wersal2015,Bernard2022}.
However, MC codes for the neutrals still have significant advantages.
\textcolor{\usecolor}{ Compared to continuum-based kinetic codes, MC
  codes apparently can more easily incorporate multiple species,
  complex atomic physics, and realistic wall geometry; the most
  detailed neutral transport models presently existing in the boundary
  plasma community are all MC-based \cite{Stotler1994,Reiter2005}.
  This motivates the search for a neutrals model that would have the
  flexibility of MC but without the statistical noise and poor
  convergence.
  }

The application of machine learning (ML) methods to the study of
physical systems is an emergent and rapidly expanding area of
research. Some of the main drivers of this rapid growth is the volume
and variety of data available today, computational power, advancement
in ML algorithms, and availability of ML software; but also the
interest to find new, innovative approaches to many existing problems.
Transport of neutral particles in plasma is such a long-standing
problem where ML methods may have promise to find more efficient
solutions.
\textcolor{\usecolor}{
Several recent papers discussed application of ML methods to neutral
transport modeling in plasma \cite{Wang2023,Zhang2025}. The
application of ML method to neutral modeling described in this report
is distinct as it is based on the concept of propagator discussed
below.  }

%%%%%%%%%%%%%%%%%%%%%%%%%%%%%%%%%%%%%%%%%%%%%%%%%%%%%%%%%%%%%%%%%%%%%%%%%%%%

%%%%%%%%%%%%%%%%%%%%%%%%%%%%%%%%%%%%%%%%%%%%%%%%%%%%%%%%%%%%%%%%%%%%%%%%%%%%
\section{Propagator model}

\subsection{MC neutral transport}

\textcolor{\usecolor}{
In the context of applications to fusion boundary plasmas, the focus
here is on the transport of hydrogen atoms in a hydrogen plasma.
}
In steady state, neglecting neutral-neutral collisions, the
distribution of neutral atoms in hydrogenic plasma obeys the kinetic
equation

\beq
v \cdot \grad f_n(v)
- n_i \int \sigma_{cx} |v-v'| f_i(v) f_n(v') dv' 
+ f_n(v) (n_i K_{cx}(v) + n_e K_i)
=
f_i(v) n_i n_e K_r
+ S_{n,ext}
\label{eq:kineq}
\eeq

Here $f_n$ is the distribution function of neutrals, $f_i$ is the
distribution function of ions, $K_r$, $K_i$, $K_{cx}$ are the rates of
recombination, ionization, and charge-exchange reactions, $S_{n,ext}$
is an external neutral source, e.g., from plasma recycling on material
surfaces.

For solving Eq.(\ref{eq:kineq}), a small 1D neutral transport code,
MC1D, was developed following the well-known standard Monte-Carlo
algorithms \cite{Heifetz1986}. For a given spatial distribution of
Maxwellian plasma profiles, $n_{e,i}(x)$ and $T_{e,i}(x)$, and for a
given neutral source, $S_n(x,v)$, the MC1D code calculates the
steady-state distribution of neutral density and other moments of the
neutral distribution function.

\subsection{Propagator}

The kinetic equation Eq. (\ref{eq:kineq}) is an inhomogeneous linear
equation for the neutral distribution function,

\beq
\hat{L} f_n = S_{n}
\label{eq:lineq}
\eeq

where $\hat{L}$ a non-local linear differential operator and
$S_{n}$ is the total neutral source.

For Eq. (\ref{eq:lineq}), one can consider the Green function (or
propagator) which is the solution of the kinetic equation with a
localized source $S_{n} \propto \delta(x-x_0)$. However, instead of
the ``full'' propagator which is the distribution function of all
neutral particles for a delta-function neutral source, it is
convenient to consider a ``single-collision'' propagator $\hat{P}$
which is the distribution function of neutral particles immediately
after their fist charge-exchange (CX) collision, for a localized neutral source. The
single-collision propagator applied to the neutral source produces the
distribution of first-generation CX neutral particles, which become a
secondary source and produce a second-generation of CX particles and
so forth. Combining CX sources from all generations, one can
calculation the total CX source, i.e., the number of CX events per
unit volume and unit time, which is simply connected to the
steady-state neutral density distribution.

Since the considered neutral transport problem is linear, for a
general neutral source, described by vector $\vec{S}_n$, one can find
the spatial distribution of ``first-generation'' CX events
$\vec{S}_{cx,1}$ as a product

\beq
\vec{S}_{cx,1} = \hat{P} \vec{S}_n
\eeq

Applying operator $\hat{P}$ to vector $\vec{S}_{cx,1}$ produces the
distribution function of ``second-generation'' CX neutrals
$\vec{S}_{cx,2}$,

\beq
\vec{S}_{cx,2} = \hat{P} \vec{S}_{cx,1} =  \hat{P}^2 \vec{S}_n
\eeq

and so forth.

Thus, the total CX source is

\beqar
\vec{S}_{cx,tot} = \vec{S}_{cx,1} + \vec{S}_{cx,2} + \vec{S}_{cx,3 } + \ldots =
(\hat{P} + \hat{P}^2 + \hat{P}^3 + \ldots) \vec{S}_n
\eeqar

For non-zero ionization rate, the norm of linear operator $\hat{P}$
is less than 1, which guarantees the convergence of the power
series. Using the identities in the Appendix, one can write the linear
equation for the CX source,

\beq
(\hat{I} - \hat{P}) \vec{S}_{cx,tot} = \hat{P} \vec{S}_n,
\eeq

where $\hat{I}$ is the identity operator.

From $\vec{S}_{cx,tot}$ one can immediately find the neutral density from

\beq
n_N = S_{cx,tot}/(n_i K_{cx},)
\eeq

where $K_{cx}$ is the rate of CX collisions, a functions of plasma
density and temperature.

\subsection{Numerical calculations}

\textcolor{\usecolor}{
For numerical experiments discussed further in this report, plasma
recombination is neglected, the rate of ionization $K_{iz}=\left<\sigma
v\right>_{iz}$ is taken constant, independent of density and temperature,
and the CX cross-section $\sigma_{cx}$ is taken constant as well.}

\textcolor{\usecolor}{
For calculation of the single-collision propagator $\hat{P}$, the MC1D
code is set up with specular reflection boundary conditions. A uniform
grid ${x_i}$, $0 \leq i \leq N-1$ covers the spatial domain, and all
spatial profiles on the grid are represented by grid functions which
are vectors of dimension $N$. Then the propagator is represented by a
matrix of dimensions $[N,N]$. The matrix elements $P_{ij}$ is the
probability for a particle originating in grid cell $i$ to suffer the
first CX collision in grid cell $j$.  For calculation of matrix
elements $P_{ij}$, a localized neutral source $\delta(x-x_i)$ is set
in the middle of $i_{th}$ grid cell, and the particles are followed
with MC1D to their first collision which can be either a CX or ionization (IZ)
event in some grid cell $j$.  Then, using an ensemble of such MC
particles, the statistical distribution (a histogram) is calculated
which allows to calculate the single-collision propagator matrix $P$.
}

\textcolor{\usecolor}{
To demonstrate the performance of the propagator approach,
Fig. (\ref{fig:prop_test_nvar},\ref{fig:prop_test_tvar}) show results
of several illustrative calculations. Fig. (\ref{fig:prop_test_nvar})
shows the results for non-uniform tanh-like plasma density profile and
uniform plasma temperature profile, and
Fig. (\ref{fig:prop_test_tvar}) shows the results for non-uniform
tanh-like plasma temperature profile and uniform plasma density
profile. In each Figure, the results are shown on the right for the
corresponding steady-state neutral density profile, for high, medium,
and low plasma collisionality set by the charge-exchange and
ionization rates shown in the plots. Results for the neutral density
are shown by the cross marks for direct MC calculation and by the blue
line for the propagator. Note that the notation ``P1'' in the plots
stands for the single-collision propagator and ``PF'' stands for the
full propagator, i.e., the direct MC method.  }

\textcolor{\usecolor}{
Clearly, for all analyzed cases in Figs. (\ref{fig:prop_test_nvar},
\ref{fig:prop_test_tvar}) the results from the single-collision
propagator approach are consistent with the direct MC results, for all
collisionality regimes. It is important to note that the
propagator-based calculation (dominated by the cost of calculation of
the propagator itself) is usually faster and more accurate than the
direct MC calculation. Detailed analysis of this will be covered in a
separate paper \cite{Parker2025}.
}
\newpage
%%%%%%%%%%%%%%%%%%%%%%%%%%%%%%%%%%%%%%%%%%%%%%%%%%%%%%%%%%%%%%%%%%%%%%%%%%%%

%%%%%%%%%%%%%%%%%%%%%%%%%%%%%%%%%%%%%%%%%%%%%%%%%%%%%%%%%%%%%%%%%%%%%%%%%%%%
\section{Neural network approximation}

\textcolor{\usecolor}{
\subsection{Neural network setup}
Constructing the propagator solves the problem of calculating the neutral
density profile for given plasma density and temperature profiles and
a given neutral source. However, the calculation of the propagator itself
involves time-consuming kinetic calculations, and for integrated
plasma-neutral modeling a faster calculation of the propagator is
desired, which can be accomplished by a neural network (NN) algorithm.
}

For numerical experiments with the NN algorithm in this report, a
spatial domain was used with $x \in [0,10]$, covered by a uniform grid
with $N$=21 grid points.  For the rates of atomic reactions,
$K_{iz}=1$ and $\sigma_{cx}$=20 were used.
For training the NN algorithm, a dataset was prepared with a range of
piecewise linear plasma density profiles and uniform temperature
profile. A plasma density profile (which is the input information) was
represented by five parameters, corresponding to plasma density values
at five collocation points $x_i = x_{min} + i (x_{max}-x_{min})/4$,
for $i \in [0,4]$. Thus, for the NN algorithm, the input is the plasma
profile represented by the five values at the collocation points, and
the output is $N^2$ elements of the propagator matrix.

A densely connected neural network is set up using the standard
software tools, with three layers: the input layer with 5 neurons, a
hidden layer, and the output layer with $N^2$ neurons. For the results
shown in this paper, 11 neurons are used in the hidden layer with
exponential activation function, while the output layer had a linear
activation function. The regression NN model is trained using mean
square error loss function and Nesterov Adaptive Moment Estimation
(NAdam) optimization algorithm. The NN model training is performed for
10000 epochs with 10$^{-3}$ learning rate.

To train the NN model, a dataset was generated with random values for
plasma density at the collocation points. These values were uniformly
distributed within the range $[0.1, 1.0]$, with a uniform plasma
temperature of 1. It was found that a dataset with 10,000 samples was
sufficient for the NN algorithm to perform well. The algorithm was
trained on 80\% of the dataset and tested on the remaining 20\%. The
results of this testing, shown in Fig. (\ref{fig:nn_performance}),
demonstrate the NN algorithm's predictive ability, lending confidence
in its performance.

\subsection{Neural network results}
The trained NN algorithm appears to be an efficient model that allows
a fast, and reasonably accurate, calculation of the propagator matrix
for a given density profile. Then, for a given neutral source $S_n$,
the spatial distribution of neutral density $n_N(x)$ can be easily
calculated.

Results of several test cases comparing neutral density profiles
predicted by direct MC, by a directly calculated propagator, and by a
NN-predicted propagator, are demonstrated in
Fig. (\ref{fig:prop_nn_test1}) and Fig. (\ref{fig:prop_nn_test2}). In
Fig. (\ref{fig:prop_nn_test1}), the results are shown for $S_n(x)$
represented by three delta-functions, and in
Fig. (\ref{fig:prop_nn_test2}) the results are shown for a neutral source
profile of general shape, see detailed description in the Figure
captions.
\textcolor{\usecolor}{ One should note that in the described numerical
  experiments, the NN-predicted propagator based calculation of
  neutral density is faster than the direct MC method by many orders
  of magnitude.  }

\textcolor{\usecolor}{
As can be seen in Figs. (\ref{fig:prop_nn_test1}, \ref{fig:prop_nn_test2}),
the neutral density profiles based on NN-predicted propagators, are
rather accurate but not exact. One cause of the accuracy loss is the
limited predictive ability of the NN algorithm, as seen in
Fig. (\ref{fig:nn_performance}). But what is apparently more
significant, the plasma density profiles used for these test problems
cannot be in general approximated as piecewise linear like those used
to train the neural network. This apparently leads to the relatively
large errors for the tanh-like density profile case, in the right
column in Figs. (\ref{fig:prop_nn_test1},\ref{fig:prop_nn_test2}).  }

%%%%%%%%%%%%%%%%%%%%%%%%%%%%%%%%%%%%%%%%%%%%%%%%%%%%%%%%%%%%%%%%%%%%%%%%%%%%

%%%\input{results}

%%%%%%%%%%%%%%%%%%%%%%%%%%%%%%%%%%%%%%%%%%%%%%%%%%%%%%%%%%%%%%%%%%%%%%%%%%%%%%%%
\textcolor{\usecolor}{
\section{Discussion}
}

\textcolor{\usecolor}{
The propagator concept used here is closely related to the idea of
multiple generations of CX neutrals that has been long used in the
experimental edge plasma community and used in the KN1D neutral
transport code developed by LaBombard \cite{Labombard2001}. The KN1D
code essentially calculates a single-collision propagator for the
neutrals, using a deterministic solution of the kinetic Boltzmann
equation for the neutrals. In the present work, the propagator is
calculated using an MC method. For generalizing to 2D/3D, calculation
of the propagator with the MC method may have advantage in treating
complex boundary conditions and complex atomic physics. Note that for
higher dimensions, a 2D/3D kinetic code will be needed to calculate
the propagator, and compared to 1D the propagator will be represented
by a much larger matrix with a different sparsity structure (multiple
bands), but the conceptual framework would apply the
same way.
}

\textcolor{\usecolor}{
For the neural network approximation of the propagator, the parameter
space would have to be substantially larger if it had to account for
the plasma electron and ion temperature profile variation $T_{e}(x)$,
$T_{i}(x)$, and for the plasma ion drift velocity
$\vec{V_i}(x)$. Including all that information, even in 1D, could
expand to a rather large computational exercise. For this report, the
choice was made to use only the plasma density variation, on a crude
spatial grid, for neural network training, to demonstrate the approach
just conceptually. More detailed treatment for general plasma profiles
will be covered in future papers.  }

\textcolor{\usecolor}{
The neural network is investigated here having in mind coupled
plasma-neutral calculations as the far-reaching goal for this
project. For such coupled simulations, one needs to update frequently
the neutral response as the plasma spatial profiles evolve in time. It
is expected that the cost of training the neural network will pay off
by providing a fast and accurate means for calculating the neutral
response in a coupled simulation. Based on preliminary analysis, an
important feature of the NN-predicted propagator is its continuous
and smooth dependence on the plasma parameters. This opens the
possibility for coupling the NN neutral model with plasma models using
Newton-based time integration methods. This is the subject of ongoing
research and the results will be reported in a subsequent paper.  }

\textcolor{\usecolor}{
It is instructive to compare the work presented here with the study in
Ref. \cite{Zhang2025}. Both studies apply neural networks to neutral
transport modeling with the common goal to improve the efficiency of
edge plasma transport simulations, but there are important differences
in the approach. The study in Ref. \cite{Zhang2025} assumes fixed gas
puffing/pumping locations, and then calculates the neutral source
terms, e.g., the ionization source. Essentially, the role of neural
network in Ref. \cite{Zhang2025} is approximation of the mapping
between the plasma profiles and the neutral source term profiles. In
principle, this approach could integrate the gas puffing/pumping
locations into the neural network; however, doing so would require
constructing a dataset that spans a much more extensive parameter
space. On the other hand, the propagator-based approach presented here
applies to an arbitrary gas source location. The propagator
is an object of higher dimensionality than the neutral sources, and
therefore it is much more expensive to calculate; but it pays off to
calculate the propagator because it can be applied for any neutral
source. This is similar to any other Green’s function: it
takes extra effort to calculate it, but once it is available, it can be
used for a wide range of applications. Note that the material wall
geometry (the domain boundary) is assumed fixed in both
Ref. \cite{Zhang2025} and the propagator-based approach discussed
here. This assumption is not very restrictive because in a major
magnetic fusion experimental device, the material wall geometry
typically changes only once every few years. On the other hand, the
location of the recycling neutral source (usually by far the dominant
neutral source) in a magnetically confined plasma experiment generally
follows the strike points location which usually changes whenever a new
magnetic equilibrium is used.}

\medskip
\medskip
\medskip
%%%%%%%%%%%%%%%%%%%%%%%%%%%%%%%%%%%%%%%%%%%%%%%%%%%%%%%%%%%%%%%%%%%%%%%%%%%%%%%%

%%%%%%%%%%%%%%%%%%%%%%%%%%%%%%%%%%%%%%%%%%%%%%%%%%%%%%%%%%%%%%%%%%%%%%%%%%%%%%%%
\section{Summary and conclusions}

In the present study, the single-collision propagator approach is
investigated for Monte-Carlo (MC) modeling of neutral particles
transport in plasma. The propagator (essentially the Green's function
for the neutral kinetic equation), depends on the plasma profiles, and
a neural network (NN) based approximation for the propagator provides a
fast and accurate way for calculating the moments of neutral
distribution function.

In numerical experiments, a 1D MC neutral transport code is used for
construction of the propagator for a set of plasma profiles, which is
used for training the NN model. Then the NN model is used for
prediction of the propagator for a given plasma density profile, which
allows for an efficient calculation of the neutral distribution
function as a convolution with a given neutral source profile. The
proposed approach, a propagator-based NN model for neutral transport
in plasma, has potential for generalization to higher dimensions and
efficient coupling with plasma models.

%%%%%%%%%%%%%%%%%%%%%%%%%%%%%%%%%%%%%%%%%%%%%%%%%%%%%%%%%%%%%%%%%%%%%%%%%%%%%%%%

%%%%%%%%%%%%%%%%%%%%%%%%%%%%%%%%%%%%%%%%%%%%%%%%%%%%%%%%%%%%%%%%%%%%%%%%%%%%
\section{Appendix}

For a power series involving a linear operator $\hat{A}$ and identity
operator $\hat{I}$, assuming the series converges, one can use the
identity which is the property of the Neumann series \cite{Kadison1986},

\beq
\hat{I} + \hat{A} + \hat{A}^2 + \ldots = (\hat{I}-\hat{A})^{-1}
\eeq

From that follow two other identities,

\beq
\hat{A} + \hat{A}^2 + \ldots = \hat{A} (\hat{I}-\hat{A})^{-1}
\eeq

and

\beq
\hat{A} + \hat{A}^2 + \ldots = (\hat{I}-\hat{A})^{-1} \hat{A}
\eeq

%%%%%%%%%%%%%%%%%%%%%%%%%%%%%%%%%%%%%%%%%%%%%%%%%%%%%%%%%%%%%%%%%%%%%%%%%%%%

%\begin{acknowledgement}
%This work was performed under the auspices of the U.S. Department of
%Energy, Office of Science, Office of Fusion Energy Sciences and Office
%of Advanced Scientific Computing Research by Lawrence Livermore
%National Laboratory (LLNL) under Contract DE-AC52-07NA27344, and by
%the LLNL MFE-FAST project, SCW01924. G.P. is supported by an NSF
%Mathematical Sciences Postdoctoral Research Fellowship (Award
%No. 2303102).
%\end{acknowledgement}

\medskip

{\small {\bf \noindent Data Availability Statement} Research data are not shared.}

%%The style of the following references should be used in all documents.
%%
%
%\input{references}
%
%\input{figs}

%%%%%%%%%%%%%%-bibliography-%%%%%%%%%%%%%%%
\clearpage
%%%%%%%%%%%%%%-bibliography-%%%%%%%%%%%%%%%
\newpage
%%%{\large \bf References}
\medskip
\medskip
\medskip
\bibliographystyle{unsrt}
\bibliography{mcnml_refs}

%\clearpage

\clearpage
%\usepackage[singlelinecheck=false,justification=justified]

%%%\section{Figure captions}
\bf{\large Figure captions}

\begin{figure}[h]
\caption{
Test problem with a single delta-peak for the neutral source and
spatially dependent plasma density profile. The plots on the left show
the profiles of plasma temperature density $N_{e,i}(x)$, $T_{e,i}(x)$
and neutral source $S_n(x)$. The plots on the right show the
steady-state neutral density profile for different levels of neutral
collisionality set by the shown scaling factors for charge-exchange
and ionization rates. Results for neutral density are shown with cross
marks for direct MC calculation and with the blue line for the
propagator.
}
\label{fig:prop_test_nvar}
\end{figure}

\begin{figure}[h]
\caption{
Test problem with a single delta-peak for the neutral source and
spatially dependent plasma temperature profile. The plots on the left
show the profiles of plasma temperature density $N_{e,i}(x)$,
$T_{e,i}(x)$ and neutral source $S_n(x)$. The plots on the right show
the steady-state neutral density profile for different levels of
neutral collisionality set by the shown scaling factors for
charge-exchange and ionization rates. Results for neutral density are
shown with cross marks for direct MC calculation and with the blue
line for the propagator.}
\label{fig:prop_test_tvar}
\end{figure}

\begin{figure}[h]
\caption{
The NN model for the propagator is trained on 80$\%$ of data in the
dataset and used to predict data in the remaining 20$\%$ of data. The
high correlation between the actual and predicted values, shown in the
plot, demonstrates successful performance of the NN model.
}
\label{fig:nn_performance}
\end{figure}

\begin{figure}[h]
\caption{
Test problem for $S_n(x)$ represented by three
delta-functions. Results for neutral density $n_N(x)$ are shown for
direct MC calculation (cross marks), directly calculated propagator
(blue line), and NN-predicted propagator (green line). The left plots
show the profiles of plasma temperature $T_{e,i}(x)$ and neutral
source $S_n(x)$. The second left column shows the plasma density
profile $n_{e,i}(x)$ (top) and the corresponding calculated neutral
density profile $n_N(x)$ (bottom) for flat $n_{e,i}(x)$. The second
right column shows results for a similar test but with
piecewise-linear $n_{e,i}(x)$. The right column shows results for a
similar test but with tanh-shaped $n_{e,i}(x)$.
}
\label{fig:prop_nn_test1}
\end{figure}

\begin{figure}[h]
\caption{
Test problem for $S_n(x)$ of a general shape. Results for neutral
density $n_N(x)$ are shown for direct MC calculation (cross marks),
directly calculated propagator (blue line), and NN-predicted
propagator (green line). The left plots show the profiles of plasma
temperature $T_{e,i}(x)$ and neutral source $S_n(x)$. The second left
column shows the plasma density profile $n_{e,i}(x)$ (top) and the
corresponding calculated neutral density profile $n_N(x)$ (bottom) for
flat $n_{e,i}(x)$. The second right column shows results for a similar
test but with piecewise-linear $n_{e,i}(x)$. The right column shows
results for a similar test but with tanh-shaped $n_{e,i}(x)$.
}
\label{fig:prop_nn_test2}
\end{figure}

%===============================================================================%

%%%%%%%%%%%%%%%%%%%%%%%%%%%%%%%%%%%%%%%%%%%%%%%%%%%%%%%%%%%%%%%%%%%%%%%%%%%%%%%%%%%%%%%%%%
%%\section{Figures}
\clearpage

{
\LARGE{Fig. 1}
\begin{figure}
\includegraphics[scale=0.40]{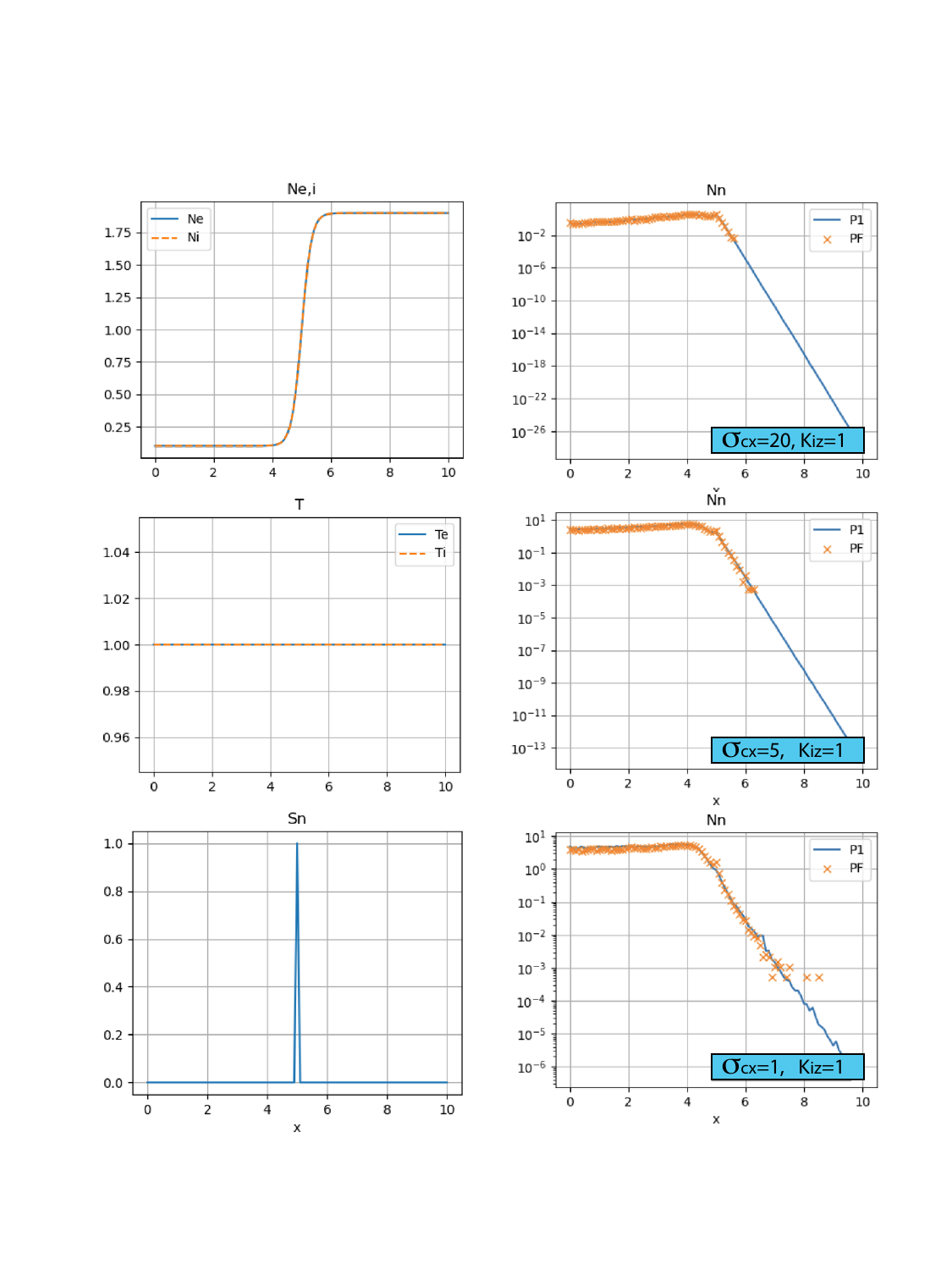}
\end{figure}
\clearpage
}

{
\LARGE{Fig. 2}
\begin{figure}
\includegraphics[scale=0.40]{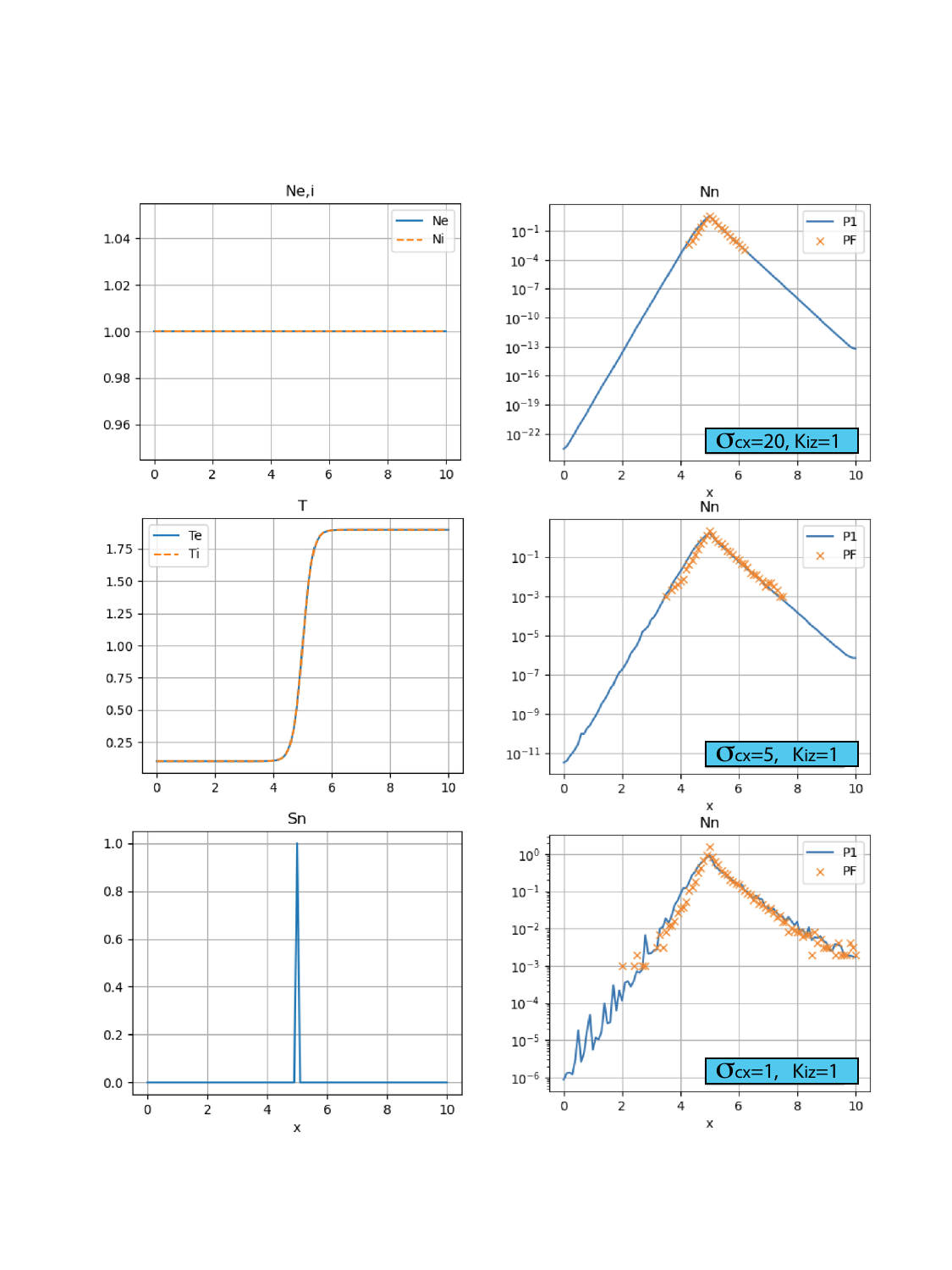}
\end{figure}
\clearpage
}

{
\LARGE{Fig. 3}
\begin{figure}
\includegraphics[scale=0.4]{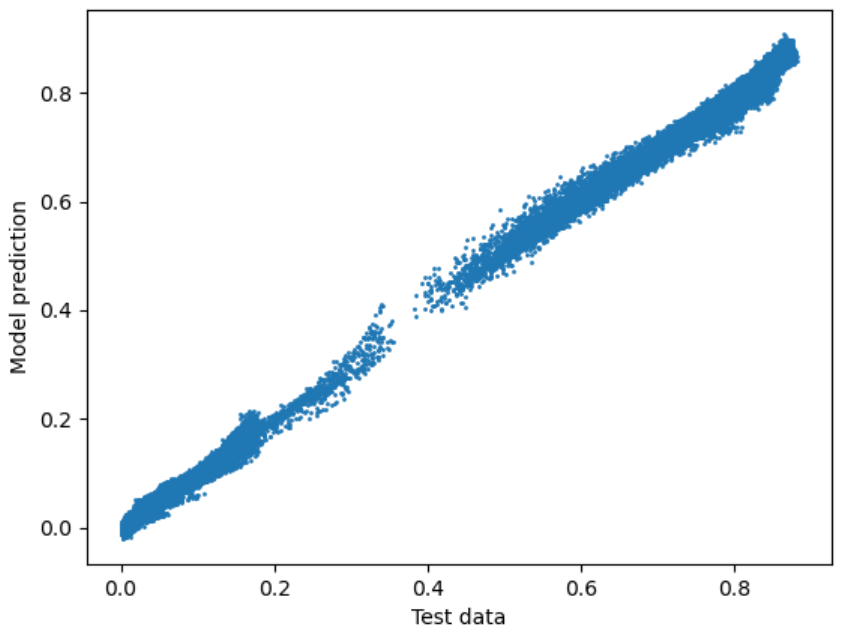}
\end{figure}
\clearpage
}

{
\LARGE{Fig. 4}
\begin{figure}
\includegraphics[scale=0.4]{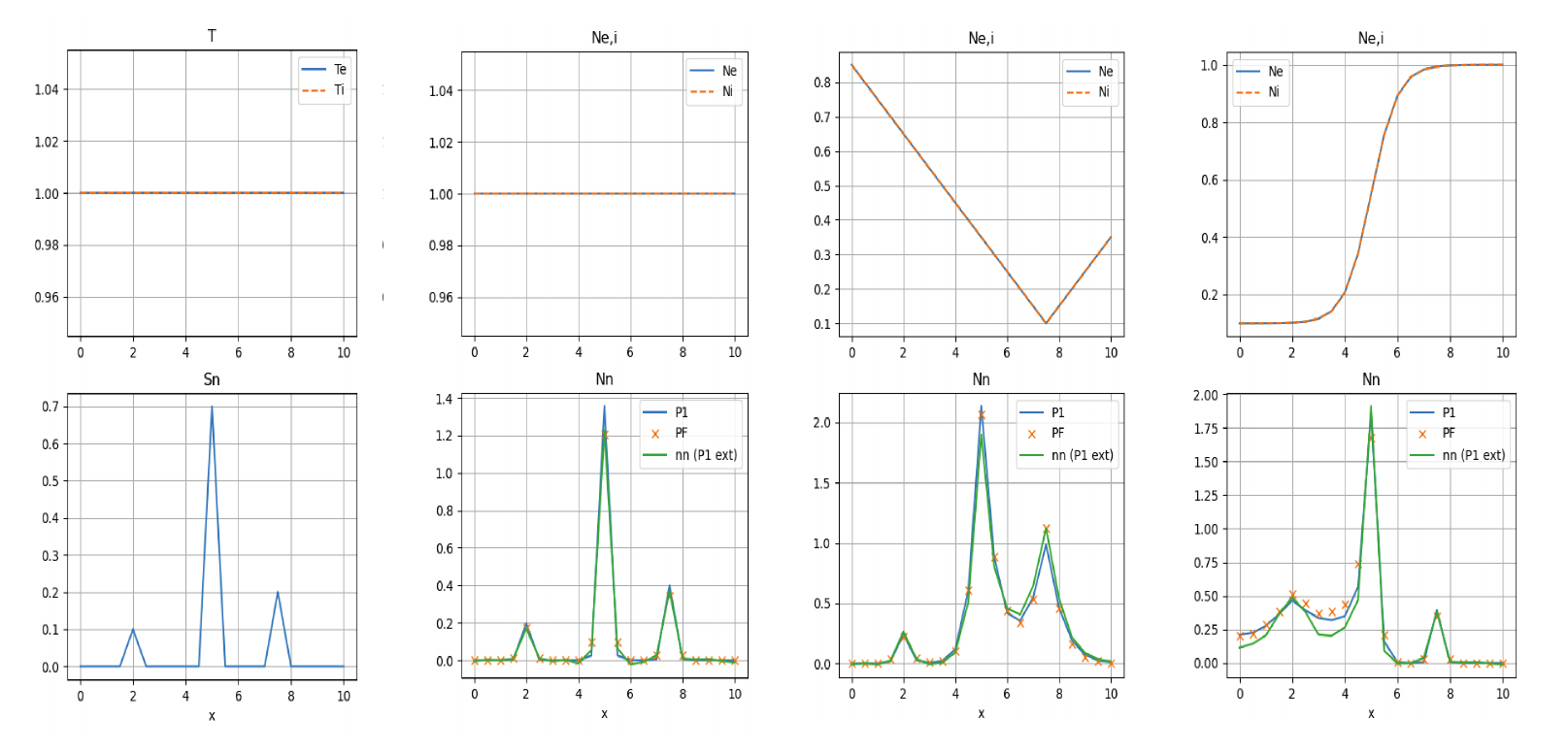}
\end{figure}
\clearpage
}

{
\LARGE{Fig. 5}
\begin{figure}
\includegraphics[scale=0.4]{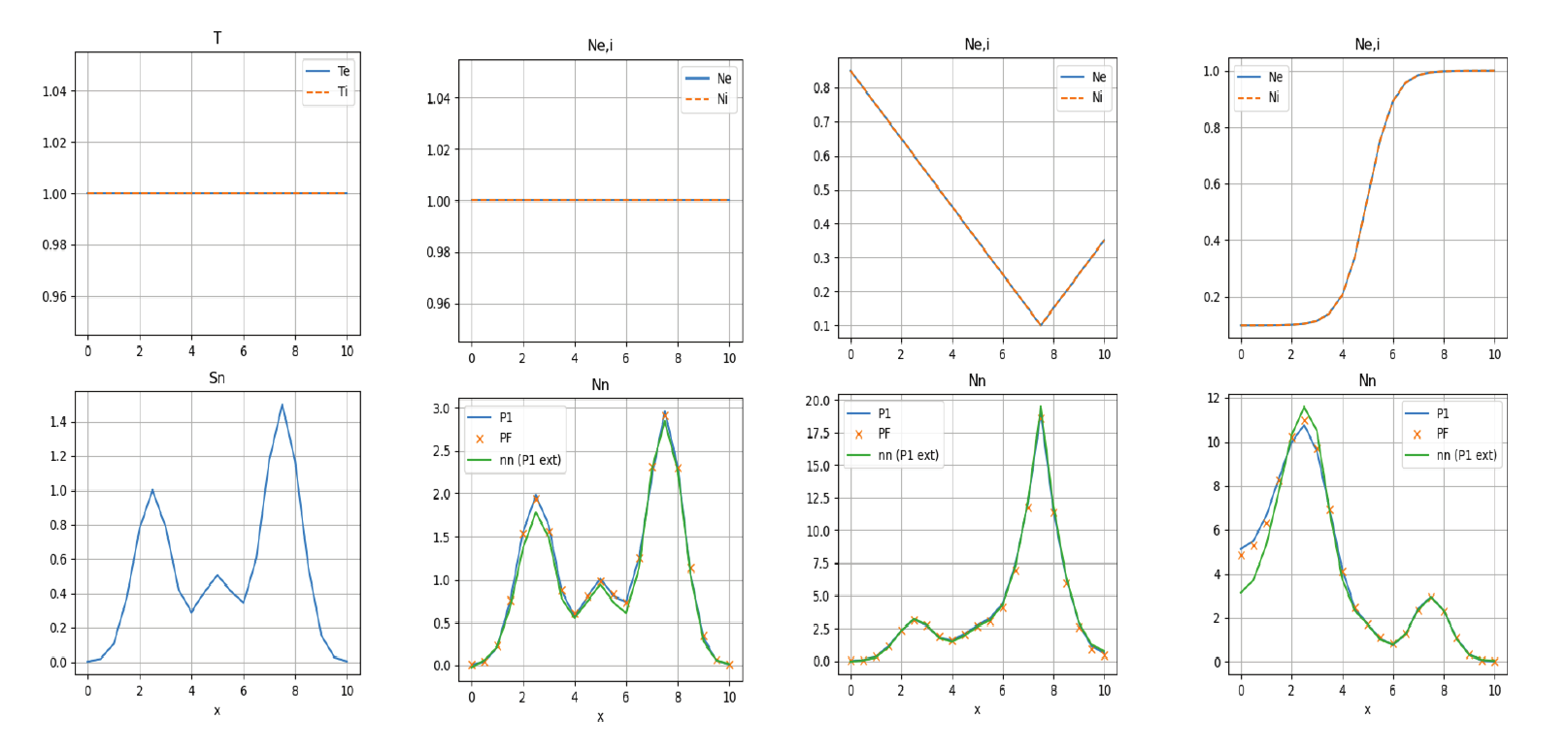}
\end{figure}
\clearpage
}

%===============================================================================%

%%%\input{figs}

\end{document}